\documentclass[]{tMOP2e}
\usepackage{axodraw,epsfig,epsf}
\usepackage{color}
\theoremstyle{plain}

\theoremstyle{remark}

\begin{document}
\title{Comparison of Electromagnetically Induced Transparency in lambda, cascade and vee three-level systems}
\author{Surajit Sen$^{\rm a}$$^{\ast}$, Tushar Kanti Dey$^{\rm a \dagger}$, Mihir Ranjan Nath$^{\rm a \ddagger}$ and Gautam Gangopadhyay$^{\rm b \P }$\thanks{$^\ast$ ssen55@yahoo.com, \; $^\dagger$ tkdey54@gmail.com, \; $^\ddagger$ mrnath\_95@rediffmail.com, \; $^\P$ gautam@bose.res.in} \\
\vspace{6pt} $^{a}$Physics Department, Guru Charan College, Silchar - 788004, India \\
 $^{b}$S N Bose National Centre for Basic Sciences,
JD Block, Sector III,
\\Salt Lake City, Kolkata 700098, India }
\maketitle
\begin{center}
\bf{Abstract}
\end{center}
\begin{abstract}
We discuss the Electromagnetically Induced Transparency (EIT) in lambda, cascade and vee type three-level systems where the Hamiltonian and the Lindblad term of each configuration are expressed in the $SU(3)$ representation. At steady state, the Optical Bloch Equations of each configuration are solved to obtain the dispersion and absorption profiles of the probe field along with their group velocities. When the EIT condition is achieved at resonance, the population oscillation shows which of the bare states are contributing to form the dark state. Our study reveals that the dark state for the lambda and cascade system effectively coincides with the lowest bare state of that system, while for the vee system, it is a maximally superposed state of the middle and upper bare states.
\end{abstract}
\vspace{1cm}
\section{Introduction}
\par
It is well known that the physical properties of the bulk material, namely, absorption coefficient, refractive index, conductivity etc can be maneuvered by the passage of light through it. In this respect, the quantum coherence plays a key role where their optical properties can be selectively modified by a number of effect, e.g., coherent population trapping (CPT)~\cite{Alzetta1976, Arimondo1976}, lasing without inversion (LWI)~\cite{Harris1989, Scully1989}, Electromagnetically Induced Transparency (EIT)~\cite{Boller1991, Harris1997,Harris1992a,Harris1992b,Marangos1996,Fleischhauer2005}, high refractive index without absorption ~\cite{Scully1991,Scully1992}, giant Kerr effect~\cite{Schmidt1996} etc. Out of them, the EIT is possibly the most successful coherent phenomenon by which the transparency of a medium can be achieved by the simultaneous action of two laser sources while satisfying some delicate conditions of the quantum interference ~\cite{Harris1997, Marangos1996, Zubairy1997}. Since its discovery in 1989~\cite{Imamoglu1989} which is followed by the experimental observation in 1990 ~\cite{Boller1991}, the study of EIT has become a promising area of research because of its amazing ability of ceasing the group velocity of light pulse when the number density of atom is very high~\cite{Hau1999, Dutton2004, Kash1999}.
\par
The manipulation of the optical property of a medium by EIT requires a clear understanding of the level structure of atoms with which the medium is composed of. In a nutshell, this phenomenon can be illustrated as follows: Two beams of coherent light, one with strong driving capability (called control or pump field) in comparison to the weaker one (called probe field), are allowed to incident on a medium which consists of three-level system. Because of an intricate role of quantum coherence at the resonant pump-probe combination, the medium, which is initially opaque to the probe field, becomes almost transparent. At steady state for the resonant probe pulse, the refractive index becomes unity and the coefficient of absorption is dropped to zero by opening a narrow transparency window called Autler-Towns splitting. More precisely, the steep and positive slope of the refractive index curve leads to the significant reduction of the group velocity of light ~\cite{Kasapi1995, Dutton2004, Hau1999}. Using the dressed atom formalism, the physics behind the transparency is explained by the trapping of all atoms into the dark state, the state which is formed due to the superposition of two mutually uncoupled bare states of the three-level system~\cite{Grey1978}.
\par
It is now widely accepted that the three-level systems, which exhibits this intriguing optical property, can be classified into the lambda, cascade and vee type of systems and they are intrinsically different from one another~\cite{Sen2012}. In literature, the theoretical study of the EIT phenomenon is extensively made primarily with the lambda system than in comparison to its other two counterparts~\cite{Welch1998, Gea-Banacloche1995,PhysRevA.83.063419}. There we have noted several contradictory remarks. For example, Ref.~\cite{Marangos1996, Gea-Banacloche1995} argue that the EIT can be observed in cascade system which contravenes with remarks of Ref.~\cite{Fleischhauer2005} that advocates the nonobservance of the EIT for the cascade and vee systems because of the absence of the meta-stable dark state. Recently, the possibility of distinct coherent process for the vee system contrast to the lambda system is hinted in some studies~\cite{PhysRevA.83.063419}. These conflicting comments on different models is possibly due to the inconspicuous and incomplete presentation of the EIT which lacks rigorous mathematical treatment of all three-level systems in equal footing. The purpose of the present work is to fill up this gap and to make a systematic study vis-$\grave{a}$-vis careful scrutiny of all three configurations to look for the underlying physical mechanism in terms of their dark state~\cite{Grey1978}.
\par
Apart from that, there is another reason for the systematic study of EIT which is known to be associated with two concomitant phenomena, namely, the coherent population trapping (CPT) and the formation of the dark state. The CPT is a standard feature of all three-level systems where whole population is being trapped in the non-evolving dark state. Although, the existence of such state is understood by noting the vanishing eigen value of the Hamiltonian of a given three-level system, but whether its formal structure remains unaltered in presence of the dissipation, is an intricate issue and, therefore, needs a closer inspection.
\par
Based on the SU(3) representation of the three-level system, recently, we have developed the model Hamiltonians of the lambda, vee and cascade systems and have solved them exactly both for the semiclassical and the quantized bichromatic fields~\cite{Sen2012}. These models form the basis of developing a rigorous and systematic treatment of EIT presented in this paper. To introduce damping using Lindlad formalism, we have used the shift operators of the SU(3) representation rather than invoking them phenomenologically. In addition, to understand the level where the atoms are being trapped, we have studied the population oscillation of each model as the function of probe-field detuning frequency. Our study reveals the constitutive bare states of the dark state for all three configurations.
\par
The remainder of the paper is organized as follows: In Section 2 we introduce the master equation of the lambda, vee and cascade systems where the model Hamiltonians and the Lindblad terms are written using the shift operators of the $SU(3)$ representation. The requisite theory to find the dispersion and absorption profiles of the probe field along with its group velocity is also outlined. Section 3 illustrates the steady state solution of the Optical Bloch Equations of the lambda, vee and cascade systems at resonant frequency of the probe field necessary to calculate the refractive index and the coefficient of absorption of the medium. Section 4 gives the numerical solution of the three systems and discuss the scenario of EIT by analyzing the population distribution in different states. Finally we summarize the main results of the paper and discuss the outlook.
\section{The models}
\par
The master equation for the three-level systems with Lindblad term is given by ($A=\Lambda,\,V,\,\Xi$),
\begin{equation}
\frac{{d\rho ^A }}{{dt}} =  \frac{i}{\hbar }\left[ {\rho^A ,H^A} \right] + \mathfrak L_D^A,
\end{equation}
\noindent
where the density matrix is defined by,
\begin{equation}
\rho^A=\left[ {\begin{array}{*{20}c}
   {\rho _{33} } & {\rho _{32} } & {\rho _{31} }  \\
   {\rho _{23} } & {\rho _{22} } & {\rho _{21} }  \\
   {\rho_{13}} & {\rho _{12} } & {\rho _{11} }  \\
\end{array}} \right],
\end{equation}
with the basis states, $|1>=(0,0,1)^T$, $|2>=(0,1,0)^T$ and $|3>=(1,0,0)^T$ being the lower, middle and upper states, respectively. The lambda, cascade and vee configurations of the three-level system are shown in Fig.1 and their levels follow same energy hierarchy condition, namely, $E_1<E_2<E_3$. In Eq.(1),  the Hamiltonian of the semiclassical three-level system in the rotating wave approximation (RWA) in the $SU(3)$ representation is  given by ~\cite{Sen2012},
\begin{subequations}
\begin{eqnarray}
H^{\Lambda} = \omega_1 V_3+\omega_2 T_3+g_{13} V_+\exp ( - i\omega _{13} t)+g _{23} T_+\exp(-i\omega_{23} t)+h.c.,
\end{eqnarray}
for the lambda system,
\begin{eqnarray}
H^\Xi = \omega_{1}U_3+\omega_{2}T_3+g_{12}U_ + \exp(-i\omega_{12}t)+g_{23}T_+\exp(-i\omega_{23} t)+h.c.,
\end{eqnarray}
for the cascade system,
\begin{eqnarray}
H^V = \omega_{1}V_3+\omega_{2}U_3+g_{13} V_ + \exp ( - i\omega _{13} t) + g_{12} U_+\exp(-i\omega_{12} t)+h.c.,
\end{eqnarray}
\end{subequations}
for the vee system, respectively. Here, $T_a, V_a, U_a$ ($a=+,-,3$) be the shift operators of $SU(3)$ representation~\cite{Greiner1994}, $g_{ij}$ ($i,j=1,2,3$) be the coupling parameters and $\omega_{ij}$ be the frequencies of the applied bi-chromatic laser fields, respectively. The $SU(3)$ Lindblad terms of three configurations appearing in Eq.(1) are given by,
\begin{subequations}
\begin{eqnarray}
\mathfrak L_D^\Lambda  = \Gamma_{31}^\Lambda (V_+V_-  \rho^\Lambda  - 2V_-  \rho^\Lambda V_+   + \rho^\Lambda V_+ V_- ) + \Gamma_{32}^\Lambda (T_+  T_-  \rho^\Lambda  - 2T_-  \rho^\Lambda T_ +   + \rho^\Lambda T_ +  T_ -),
\end{eqnarray}
for the lambda system,
{\setlength\arraycolsep{2pt}
\begin{eqnarray}
\mathfrak L_D^\Xi &=& \Gamma _{32}^\Xi(T_+  T_-  \rho^\Xi  - 2T_-  \rho^\Xi T_ +   + \rho^\Xi T_ +  T_ -) + \Gamma _{21}^\Xi(U_ +  U_ -  \rho^\Xi  - 2U_ -  \rho^\Xi U_ +  + \rho^\Xi U_ +  U_ -),
\end{eqnarray}}
for the cascade systems,
{\setlength\arraycolsep{2pt}
\begin{eqnarray}
\mathfrak L_D^V  &=& \Gamma _{31}^V(V_+V_-  \rho^V  - 2V_-  \rho^V V_+   + \rho^V V_+ V_- ) + \Gamma _{21}^V(U_ +  U_ -  \rho^V  - 2U_ -  \rho^V U_ +  + \rho^V U_ +  U_ -),
\end{eqnarray}}
\end{subequations}
for the vee system, respectively. It is evident from Fig.1 that for the lambda system, the spontaneous decay from lower to higher level ( i.e., 1 to 3 and 2 to 3) is not permitted and therefore, $\Gamma_{13}^\Lambda=\Gamma_{23}^\Lambda=0$. For the same reason, we have $\Gamma_{12}^\Xi=\Gamma_{13}^\Xi=0$ for the cascade system and $\Gamma_{13}^V=\Gamma_{12}^V=0$ for the vee system, respectively. Thus, as shown in Fig.1, the control and probe fields correspond $2\leftrightarrow 3$ and $1\leftrightarrow 3$ dipole transition for the lambda system, $1\leftrightarrow 2$ and $1\leftrightarrow 3$ transition for the vee system and $2\leftrightarrow 3$ and $1\leftrightarrow 2$ transition for the cascade system, respectively. The generic detuning condition in the $SU(3)$ representation is given by $\Delta^{A}_{kl}=(m \omega_1+ n \omega_2-\omega_{kl})$, where $m,n \in \mathbb{Z}$~\cite{Sen2012}.
\par
To understand the dispersion and absorption profiles of the medium at resonance, we need to evaluate the steady state density matrix
$\rho^A _{s}$ by solving the equation $\dot \rho^A(t)=0$ from Eq.(1) formally known as Optical Bloch Equation (OBE). Given with the steady state density matrix $\rho^A _{s}$, it is convenient to express the real and imaginary parts of the susceptibility in terms of Bloch vectors which gives the refractive index and coefficient of absorption of the probe field,
\begin{eqnarray}
n^A(\Delta^A_{ij})=1+\frac{N_0\mu_{ij}^2}{2\epsilon_0 \hbar}Tr[\rho^A _{s}\lambda_a],\\
\alpha^A(\Delta^A_{ij})=\frac{N_0\mu_{ij}^2}{2\epsilon_0 \hbar}Tr[\rho^A _{s}\lambda_{\bar a}],
\end{eqnarray}
respectively, where $N_0$ be the number of three-level configuration in the sample, $\mu_{ij}$ be the dipole moment of $i \leftrightarrow j$ transition and $\epsilon_0$ be the permittivity of free space, respectively. In Eqs.(5) and (6), the Gellmann matrix $\lambda_a$ is judiciously chosen so that it projects requisite terms from the density matrix which contributes in $i \leftrightarrow j$ dipole transition.

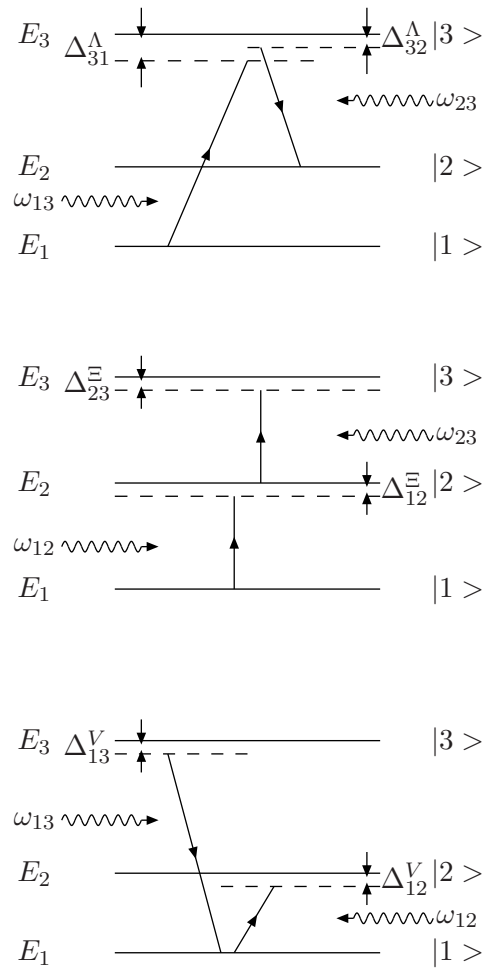
\begin{figure}[h]
\begin{center}
\begin{center}\begin{picture}(300,86)(0,0)

\Line(100,90)(200,90)
\Text(210,89)[]{$\Delta_{32}^\Lambda$}
\Text(230,90)[]{$|3>$}
\Text(70,90)[]{$E_3$}
\ArrowLine(155,85)(170,40)

\Line(100,40)(200,40)
\Text(230,40)[]{$|2>$}
\Text(70,40)[]{$E_2$}
\DashLine(100,80)(175,80){5} \DashLine(150,85)(200,85){5}
\LongArrow(110,100)(110,90) \LongArrow(110,70)(110,80)
\LongArrow(195,100)(195,90) \LongArrow(195,75)(195,85)
\ArrowLine(120,10)(150,80)
\Line(100,10)(200,10)
\Text(230,10)[]{$|1>$}
\Text(70,10)[]{$E_1$}
\Text(90,85)[]{$\Delta_{31}^\Lambda$}

\Photon(80,27)(110,27){2}{6}
\LongArrow(110,27)(115,27)
\Text(70,27)[]{$\omega_{13}$}

\Photon(190,65)(220,65){2}{6}
\LongArrow(190,65)(185,65)
\Text(230,65)[]{$\omega_{23}$}
\end{picture} \\
\end{center}
\vspace{1.5cm}
\begin{picture}(300,86)(0,0)

\Line(100,90)(200,90)
\Text(230,90)[]{$|3>$}
\Text(70,90)[]{$E_3$}
\ArrowLine(155,50)(155,85)
\Line(100,50)(200,50)
\Text(70,50)[]{$E_2$}
\Text(230,50)[]{$|2>$}
\DashLine(100,85)(200,85){5}
\DashLine(100,45)(200,45){5}
\ArrowLine(145,10)(145,45)
\Line(100,10)(200,10)
\Text(70,10)[]{$E_1$}
\Text(230,10)[]{$|1>$}
\LongArrow(110,98)(110,90) \LongArrow(110,78)(110,85)
\LongArrow(195,58)(195,50) \LongArrow(195,38)(195,45)
\Text(90,88)[]{$\Delta_{23}^\Xi$}
\Text(210,48)[]{$\Delta_{12}^\Xi$}

\Photon(80,26)(110,26){2}{6}
\LongArrow(110,26)(115,26)
\Text(70,26)[]{$\omega_{12}$}

\Photon(190,68)(220,68){2}{6}
\LongArrow(190,68)(185,68)
\Text(230,68)[]{$\omega_{23}$}
\end{picture} \\
\vspace{1.5cm}
\begin{center}\begin{picture}(300,86)(0,0)

\Line(100,90)(200,90)
\Text(230,90)[]{$|3>$}
\Text(70,90)[]{$E_3$}
\ArrowLine(120,85)(140,10)
\Line(100,40)(200,40)
\Text(230,40)[]{$|2>$}
\Text(70,40)[]{$E_2$}
\DashLine(100,85)(150,85){5}
\DashLine(140,35)(200,35){5}
\ArrowLine(145,10)(160,35)
\Line(100,10)(200,10)
\Text(230,10)[]{$|1>$}
\Text(70,10)[]{$E_1$}
\LongArrow(110,98)(110,90) \LongArrow(110,78)(110,85)
\LongArrow(195,48)(195,40) \LongArrow(195,28)(195,35)
\Text(90,88)[]{$\Delta_{13}^V$}
\Text(210,38)[]{$\Delta_{12}^V$}

\Photon(80,60)(110,60){2}{6}
\LongArrow(110,60)(115,60)
\Text(70,60)[]{$\omega_{13}$}

\Photon(190,23)(220,23){2}{6}
\LongArrow(190,23)(185,23)
\Text(230,23)[]{$\omega_{12}$}
\end{picture} \\
\vspace{1.5cm}
\end{center}
\end{center}
\caption{Lambda, Cascade and Vee type three-level configurations with the energy levels arranged as $E_3>E_2>E_1$. The detuning from the applied field are  $\Delta^{\Lambda}_{23}=\omega_1+2\omega_2-\omega_{23}$ and $\Delta^{\Lambda}_{13}=2\omega_1+ \omega_2-\omega_{13}$ for the lambda system, $\Delta^{\Xi}_{23}=-\omega_1+2\omega_2-\omega_{23}$ and $\Delta^{\Xi}_{12}=2 \omega_1- \omega_2-\omega_{12}$ for the cascade system and $\Delta^{V}_{12}= 2\omega_1+ \omega_2-\omega_{12}$ and $\Delta^{V}_{13}=\omega_1+2\omega_2-\omega_{13}$ for the vee system, respectively ~\cite{Sen2012}. \label{}}
\end{figure}
\par
Finally, the group velocity of probe field through the medium is calculated using $v_g=c/n_g$, where the group refractive index for the probe field of frequency $\omega_{ij}$ is given by,
\begin{eqnarray}
n^A_g(\Delta^A_{ij})&=& n^A(\Delta^A_{ij})+\omega_{ij}\frac{d n^A(\Delta^A_{ij})}{d\omega_{ij}}\nonumber\\
&=& n^A(\Delta^A_{ij})+\omega_{ij}\frac{\partial n^A(\Delta^A_{ij})}{\partial \big|\Delta^A_{ij}\big|}\cdot \frac{\partial \big|\Delta^A_{ij}\big|}{\partial\omega_{ij}},
\end{eqnarray}
By noting the fact that, $\frac{d\big|\Delta^A_{kl}\big|}{d\omega_{kl}}=1$, the group velocity is simplified to,
\begin{equation}
v^A_g(\Delta^A_{ij})= c/\big(1+\frac{N_0\mu_{ij}^2\omega_{ij}}{2\epsilon_0 \hbar}\frac{\partial(Tr[\rho_{s}^A\lambda_a])}{\partial \Delta^A_{ij}}\big).
\end{equation}
The positive slope of the dispersion curve near resonance ensures the reduction of the group velocity which is different for different three-level systems. In the subsequent Sections we proceed to solve the master equations with the aforesaid $SU(3)$ Lindblad terms to find steady state density matrix of all three configurations and explore the EIT effect and its associated phenomena, e.g., the group velocity reduction, coherent trapping, dark state etc.
\section{EIT in three-level systems}
\subsection{ Lambda system}
\par
To obtain the Optical Bloch Equation of the lambda system we consider the master equation  with the Hamiltonian given by Eq.(3a) and the corresponding Lindblad term in Eq.(4a). The following substitutions eliminate the phase terms  appearing in the density matrix equations,
{\setlength\arraycolsep{2pt}
\begin{eqnarray}
\rho _{11}^\Lambda  & \rightarrow & \tilde \rho_{11}^\Lambda = \rho_{11}^\Lambda,\nonumber\\
\rho _{22}^\Lambda & \rightarrow & \tilde \rho _{22}^\Lambda = \rho_{22}^\Lambda,\nonumber\\
\rho _{33}^\Lambda  & \rightarrow & \tilde \rho _{33}^\Lambda = \rho_{33}^\Lambda,\nonumber\\
\rho _{13}^\Lambda   & \rightarrow &  \tilde \rho _{13}^\Lambda  =  \tilde{\rho}_{31}^{\Lambda *} =  e^{ i\omega _{13} t} \rho _{13}^\Lambda,\nonumber \\
\rho _{12}^\Lambda    & \rightarrow & \tilde \rho _{12}^\Lambda  = \tilde{\rho}_{21}^{\Lambda *} = e^{ i(\omega _{13}  - \omega _{23} )t} \rho _{12}^\Lambda,\nonumber \\
\rho _{23}^\Lambda  & \rightarrow & \tilde \rho _{23}^\Lambda  = \tilde{\rho}_{32}^{\Lambda *} = e^{ i\omega _{23} t} \rho _{23}^\Lambda.
\end{eqnarray}}
\noindent
Using the detuning conditions to be, $\Delta _{23}^\Lambda = \omega _1  + 2\omega _2  -\omega _{23}  $ and $\Delta^\Lambda_{13} = 2\omega _1  + \omega _2 - \omega _{13} $, respectively, we obtain the following OBE (here we drop the tilde sign),
{\setlength\arraycolsep{2pt}
\begin{eqnarray}
\dot \rho _{11}^{\Lambda}  &=&  i g_{13} (\rho_{13}^\Lambda-\rho_{31}^\Lambda)+2 \Gamma_{31}^\Lambda\rho_{33}^\Lambda \nonumber\\
\dot \rho _{22}^{\Lambda}  &=& i g_{23} (\rho_{23}^\Lambda-\rho_{32}^\Lambda)+2 \Gamma_{32}^\Lambda\rho_{33}^\Lambda \nonumber\\
\dot \rho _{33}^{\Lambda}  &=& -i g_{13} (\rho_{13}^\Lambda-\rho_{31}^\Lambda)-i g_{23}(\rho_{23}^\Lambda-\rho_{32}^\Lambda)-2 (\Gamma_{31}^\Lambda+\Gamma_{32}^\Lambda) \rho_{33}^\Lambda \nonumber \\
\dot \rho _{12}^{\Lambda}  &=& \dot \rho _{21}^{\Lambda*}  = i (\Delta_{13}^\Lambda \rho_{12}^\Lambda-\Delta_{23}^\Lambda \rho_{12}^\Lambda+ g_{23} \rho_{13}^\Lambda-g_{13} \rho_{32}^\Lambda) \nonumber \\
\dot \rho _{13}^{\Lambda}  &=& \dot \rho _{31}^{\Lambda*}  = i (g_{23} \rho_{12}^\Lambda+(i (\Gamma_{31}^\Lambda+\Gamma_{32}^\Lambda)+\Delta_{13}^\Lambda) \rho_{13}^\Lambda+g_{13}
          (\rho_{11}^\Lambda-\rho_{33}^\Lambda)) \nonumber \\
\dot \rho_{23}^{\Lambda}  &=& \dot \rho _{32}^{\Lambda*}  = i( g_{13} \rho_{21}^\Lambda+  \Delta_{23}^\Lambda \rho_{23}^\Lambda) -(\Gamma_{31}^\Lambda+\Gamma_{32}^\Lambda) \rho_{23}^\Lambda+g_{23}
       (\rho_{22}^\Lambda-\rho_{33}^\Lambda)).
\end{eqnarray}}
Finally, taking the pump detuning vanishing, i.e., $\Delta_{23}^\Lambda=0$, and the normalization condition, $Tr[\rho^\Lambda]=\rho^\Lambda_{11}+\rho^\Lambda_{22}+\rho^\Lambda_{33}=1$, the steady state solutions $\rho^\Lambda_s$ of the OBE of the lambda system, namely $\dot \rho^\Lambda(t)=0$, gives,
\begin{subequations}
{\setlength\arraycolsep{2pt}
\begin{eqnarray}
{\rho_{11s}^{\Lambda}}^{} &=&  \frac{{\rho_{11N}^{\Lambda}}}{D^\Lambda}, \quad {\rho_{22s}^{\Lambda}}^{} =  \frac{{\rho_{22N}^{\Lambda}}}{D^\Lambda},\quad {\rho_{33s}^{\Lambda}}^{} =  \frac{{\rho_{33N}^{\Lambda}}}{D^\Lambda}\\
{\rho_{12s}^{\Lambda}} ={{\rho_{21s}^{\Lambda*}}}& = &\frac{{\rho_{12N}^{\Lambda}}}{D^\Lambda}, \quad {\rho_{13s}^{\Lambda}}^{}  = {{\rho_{31s}^{\Lambda}}}^{*}=  \frac{{\rho_{13N}^{\Lambda}}}{D^\Lambda},\quad {\rho_{23s}^{\Lambda}}^{} ={{\rho_{32s}^{\Lambda*}}} =  \frac{{\rho_{23N}^{\Lambda}}}{D^\Lambda},
\end{eqnarray}}
\end{subequations}
\noindent where the numerator and the denominators are given in Appendix. The steady state density matrix $\rho^\Lambda_{s}$ can be now used to calculate the refractive index, coefficient of absorption and the group velocity of the lambda system from Eqs.(5), (6) and (7),
\begin{subequations}
\begin{eqnarray}
n^{\Lambda}_{13}(\Delta^{\Lambda}_{13})&=&1+\frac{N_0\mu_{13}^2}{2\epsilon_0 \hbar}Tr[\rho^{\Lambda}_{s}\lambda_4],\\
\alpha^{\Lambda}_{13}(\Delta^{\Lambda}_{13})&=&\frac{N_0\mu_{13}^2}{2\epsilon_0 \hbar}Tr[\rho^{\Lambda}_{s}\lambda_5],\\
v^{\Lambda}_g(\Delta^{\Lambda}_{13})&=&{c}/\left(1+\frac{N_0\mu_{13}^2\omega_{13}}{2\epsilon_0 \hbar}\frac{\partial(Tr[\rho^{\Lambda}_{s}\lambda_4])}{\partial \Delta_{13}}\right),
\end{eqnarray}
\end{subequations}
respectively. We note that $\lambda_4$ and $\lambda_5$ appearing in Eqs.(12) has allowed the $1\leftrightarrow 3$ dipole transition appeared in the dispersion and absorption profiles of the probe field.
\subsection{ Cascade system}
Proceeding in the similar way for the cascade system we make the following substitutions,
{\setlength\arraycolsep{2pt}
\begin{eqnarray}
\rho _{11}^{\Xi}  & \rightarrow & \tilde \rho_{11}^{\Xi} = \rho _{11}^{\Xi}, \nonumber \\
\rho _{22}^{\Xi}  & \rightarrow & \tilde \rho _{22}^{\Xi} = \rho _{22}^{\Xi},\nonumber\\
\rho _{33}^{\Xi}  & \rightarrow & \tilde \rho _{33}^{\Xi} = \rho _{33}^{\Xi},\nonumber\\
\rho _{12}^{\Xi}  & \rightarrow & \tilde \rho _{12}^{\Xi} =  \tilde{\rho}_{21}^{\Xi *} = e^{ i\omega _{12} t} \rho _{12}^{\Xi},\nonumber \\
\rho _{23}^{\Xi}  & \rightarrow & \tilde \rho _{23}^{\Xi} = \tilde{\rho}_{32}^{\Xi *}  = e^{ i\omega _{23} t} \rho _{23}^{\Xi},\nonumber \\
\rho _{13}^{\Xi}  & \rightarrow & \tilde \rho _{13}^{\Xi} = \tilde{\rho}_{31}^{\Xi *} = e^{ i(\omega _{12}+\omega _{23} )t} \rho _{13}^{\Xi},
\end{eqnarray}}
and taking the detuning conditions to be, $\Delta^{\Xi}_{12} = 2\omega _1  - \omega _2  -\omega _{12}$ and $\Delta^{\Xi}_{23} = 2\omega _2  - \omega _1 - \omega _{23} $, we obtain the OBE for the cascade system,
{\setlength\arraycolsep{2pt}
\begin{eqnarray}
\dot \rho _{11}^{\Xi}  &=&i g_{12} (\rho_{12}^{\Xi}-\rho_{21}^{\Xi})+2\Gamma_{21}^{\Xi} \rho_{22}^{\Xi} \nonumber \\
\dot \rho _{22}^{\Xi}  &=& -i (g_{12} (\rho_{12}^{\Xi}-\rho_{21}^{\Xi})+g_{23} (\rho_{32}^{\Xi}-\rho_{23}^{\Xi}))+2 \Gamma_{32}^{\Xi}\rho_{33}^{\Xi} -2 \Gamma_{21}^{\Xi} \rho_{22}^{\Xi}\nonumber \\
\dot \rho _{33}^{\Xi}  &=& i (g_{23}\rho_{32}^{\Xi}-g_{23}\rho_{23}^{\Xi})+2 \Gamma_{32}^{\Xi}\rho_{33}^{\Xi} \nonumber \\
\dot \rho _{12}^{\Xi}  &=&i (g_{23}\rho_{12}^{\Xi}+\Delta_{12}^{\Xi}\rho_{13}^{\Xi}+\Delta_{23}^{\Xi}\rho_{13}^{\Xi}-g_{12}\rho_{23}^{\Xi})-\Gamma_{32}^{\Xi} \rho_{13}^{\Xi}\nonumber \\
\dot \rho _{13}^{\Xi}  &=&i (g_{23}\rho_{12}^{\Xi}+\Delta_{12}^{\Xi}\rho_{13}^{\Xi}+\Delta_{23}^{\Xi}\rho_{13}^{\Xi}-g_{12} \rho_{23}^{\Xi})-\Gamma_{32}^{\Xi} \rho_{13}^{\Xi}\nonumber \\
\dot \rho _{23}^{\Xi}  &=&i (-g_{12}\rho_{13}^{\Xi}+g_{23}\rho_{22}^{\Xi}+\Delta_{23}^{\Xi} \rho_{23}^{\Xi}-g_{23}\rho_{33}^{\Xi})-\Gamma_{21}^{\Xi} \rho_{23}^{\Xi}-\Gamma_{32}^{\Xi}\rho_{23}^{\Xi}
\end{eqnarray}}
Finally choosing $\Delta_{23}^{\Xi}=0$ and taking the normalization to be, $Tr[\rho^{\Xi}]=1$, the steady state solution of the density matrix of the cascade system is given by,
\begin{subequations}
{\setlength\arraycolsep{2pt}
\begin{eqnarray}
 {\rho_{11s}^{\Xi}}^{} &=&  \frac{{\rho_{11N}^{\Xi}}}{D^{\Xi}}, \quad {\rho_{22s}^{\Xi}}^{} = \frac{{\rho_{22N}^{\Xi}}}{D^{\Xi}}, \quad{\rho_{33s}^{\Xi}}^{} =  \frac{{\rho_{33N}^{\Xi}}}{D^{\Xi}},\\
\quad {\rho_{12s}^{\Xi}} &=& {{\rho_{21s}^{\Xi*}}} = \frac{{\rho_{12N}^{\Xi}}}{D^{\Xi}},\quad {\rho_{13s}^{\Xi}}^{} = {{\rho_{31s}^{\Xi*}}}=  \frac{{\rho_{13N}^{\Xi}}}{D^{\Xi}},\quad {\rho_{23s}^{\Xi}}^{} ={{\rho_{32}^{\Xi*}}} =  \frac{{\rho_{23N}^{\Xi}}}{D^{\Xi}},
\end{eqnarray}}
\end{subequations}
where the elements of the cascade system are given in Appendix. The steady state density matrix gives the refractive index, coefficient of absorption and the group velocity of the cascade system,
\begin{subequations}
\begin{eqnarray}
n^{\Xi}_{12}(\Delta_{12}^{\Xi})&=&1+\frac{N_0\mu_{12}^2}{2\epsilon_0 \hbar}Tr[\rho^{\Xi}_{s}\lambda_6],\\
\alpha^{\Xi}_{12}(\Delta_{12}^{\Xi})&=&\frac{N_0\mu_{12}^2}{2\epsilon_0 \hbar}Tr[\rho^{\Xi}_{s}\lambda_7],\\
v^{\Xi}_g(\Delta_{12}^{\Xi})&=&{c}/\left(1+\frac{N_0\mu_{12}^2\omega_{12}}{2\epsilon_0 \hbar}\frac{\partial(Tr[\rho^{\Xi}_{s}\lambda_6])}{\partial \Delta_{12}}\right),
\end{eqnarray}
\end{subequations}
respectively. We note that $\lambda_6$ and $\lambda_7$ matrices appearing in the refraction and absorption profiles in Eq.(16)
corresponds to $1\leftrightarrow 2$ dipole transitions of the probe field.
\subsection{\label{1x} Vee system}
To obtain the Optical Bloch Equation for the vee system we consider the following substitutions,
{\setlength\arraycolsep{2pt}
\begin{eqnarray}
\rho _{11}^V  & \rightarrow & \tilde \rho_{11}^V = \rho _{11}^V,\nonumber\\
\rho _{22}^V  & \rightarrow & \tilde \rho _{22}^V= \rho _{22}^V,\nonumber\\
\rho _{33}^V  & \rightarrow & \tilde \rho _{33}^V = \rho _{33}^V,\nonumber\\
\rho _{13}^V  & \rightarrow & \tilde \rho _{13}^V =  \tilde{\rho}_{31}^{V *} = e^{ i\omega _{13} t} \rho _{13}^V,\nonumber \\
\rho _{23}^V  & \rightarrow & \tilde \rho _{23}^V =  \tilde{\rho}_{32}^{V *}  = e^{ i(\omega _{13}  - \omega _{12} )t} \rho _{23}^V,\nonumber \\
\rho _{12}^V  & \rightarrow & \tilde \rho _{12}^V =  \tilde{\rho}_{21}^{V *} = e^{ i\omega _{12} t} \rho _{12}^V.
\end{eqnarray}}
Using the detuning conditions, $\Delta^V_{13} = \omega _1  + 2\omega _2  -\omega _{13}  $ and $\Delta^V _{12} = 2\omega _1  + \omega _2 - \omega _{12} $, we obtain the following OBE,
{\setlength\arraycolsep{2pt}
\begin{eqnarray}
\dot \rho _{11}^{{\text{V}}}  &=& 2\Gamma_{21}^{\text{V}} \rho_{22}^{\text{V}}+i (g_{12} (\rho_{12}^{\text{V}}-\rho_{21}^{\text{V}})+g_{13} (\rho_{13}^{\text{V}}-\rho_{31}^{\text{V}})+2 \Gamma_{31}^{\text{V}}\rho_{33}^{\text{V}} \nonumber\\
\dot \rho _{22}^{{\text{V}}}  &=& -i g_{12} (\rho_{12}^{\text{V}}-\rho_{21}^{\text{V}})-2 \Gamma_{21}^{\text{V}}\rho_{22}^{\text{V}} \nonumber\\
\dot \rho _{33}^{{\text{V}}}  &=& -i g_{13} (\rho_{13}^{\text{V}}-\rho_{31}^{\text{V}})-2 \Gamma_{31}^{\text{V}}\rho_{33}^{\text{V}} \nonumber\\
\dot \rho _{12}^{{\text{V}}}  &=& \dot \rho _{21}^{{\text{V}}*}  =i (g_{12} \rho_{11}^{\text{V}}+\Delta_{12}^{\text{V}} \rho_{12}^{\text{V}}-g_{12} \rho_{22}^{\text{V}}-g_{13} \rho_{32}^V)-\Gamma_{21}^{\text{V}} \rho_{12}^{\text{V}} \nonumber\\
\dot \rho _{13}^{{\text{V}}}  &=& \dot \rho_{31}^{{\text{V}}*}  =i (g_{13} \rho_{11}^{\text{V}}+\Delta_{13}^{\text{V}} \rho_{13}^{\text{V}}-g_{12} \rho_{23}^{\text{V}}-g_{13} \rho_{33}^{\text{V}})-\Gamma_{31}^{\text{V}}\rho_{13}^{\text{V}} \nonumber\\
\dot \rho _{23}^{{\text{V}}}  &=& \dot \rho _{32}^{{\text{V}}*}  = -\Gamma_{21}^{\text{V}}\rho_{23}^{\text{V}}-\Gamma_{31}^{\text{V}} \rho_{23}^{\text{V}}+i (-g_{12}\rho_{13}^{\text{V}}+g_{13}\rho_{21}^{\text{V}}-\Delta_{12}^{\text{V}}\rho_{23}^{\text{V}}+\Delta_{13}^{\text{V}}
   \rho_{23}^{\text{V}}).
\end{eqnarray}}
Once again, taking the pump detuning to be $\Delta_{12}^V=0$ and $Tr[\rho^V]=1$, the steady state solution of the above coupled equations are given by,
\begin{subequations}
{\setlength\arraycolsep{2pt}
\begin{eqnarray}
{\rho_{11s}^{V}}^{} &=&  \frac{{\rho_{11N}^{V}}}{D^V}, \quad {\rho_{22s}^{V}}^{} = \frac{{\rho_{22N}^{V}}}{D^V},\quad {\rho_{33s}^{V}}^{} =  \frac{{\rho_{33N}^{V}}}{D^V}\\
{\rho_{12s}^{V}} &=& {{\rho_{21s}^{V}}}^{*} = \frac{{\rho_{12N}^{V}}}{D^V},\quad
{\rho_{13s}^{V}}^{} ={{\rho_{31s}^{V}}}^{*}=  \frac{{\rho_{13N}^{V}}}{D^V},\quad {\rho_{23s}^{V}}^{} ={{\rho_{32s}^{V}}}^{*} =  \frac{{\rho_{23N}^{V}}}{D^V},
\end{eqnarray}}
\end{subequations}
\noindent
respectively, where the numerators and the denominators of the vee system are presented in Appendix. Similar to the previous case, at the steady state, the refractive index, coefficient of absorption and the group velocity of the vee system can be easily calculated,
\begin{subequations}
\begin{eqnarray}
n^V_{13}(\Delta_{13}^{\text{V}})&=&1+\frac{N_0\mu_{13}^2}{2\epsilon_0 \hbar}Tr[\rho^V_{s}\lambda_4],\\
\alpha^V_{13}(\Delta_{13}^{\text{V}})&=&\frac{N_0\mu_{13}^2}{2\epsilon_0 \hbar}Tr[\rho^V_{s}\lambda_5],\\
v^V_g(\Delta_{13}^{\text{V}})&=&{c}/\left(1+\frac{N_0\mu_{13}^2\omega_{13}}{2\epsilon_0 \hbar}\frac{\partial(Tr[\rho^{V}_{s}\lambda_4])}{\partial \Delta_{13}}\right).
\end{eqnarray}
\end{subequations}
We note that, similar to the lambda system, $\lambda_4$ and $\lambda_5$ matrices appearing in the refraction and absorption profiles in Eq.(20) correspond to the $1\leftrightarrow 3$ dipole transition. Having developing the basic tenets of the EIT theory, we now proceed for the numerical results to illustrate its physical content.
\section{\label{4} Numerical results}
\par
We now compare the EIT exhibited by the lambda, cascade and vee configurations by analyzing their steady state population dynamics with special emphasis on a succinct phenomenon of the formation of dark state. To reveal it quantitatively, we take the following numerical values: let $N_0=10^{21}$ be the value of the number of three-level system of given configuration per unit cubic-meter and $\mu_{ij} (i \neq j)$ be the dipole moment which is approximated by the Bohr magneton. The control beam frequency and the Rabi frequency parameter are chosen to be $\omega^\Lambda_{13}=2.37\times10^{9}$ MHz and $g_{23}=105$ MHz for the lambda system ~\cite{PhysRevA.51.R2703}, $\omega^\Xi_{13}=2.88\times10^{9}$ MHz and $g_{23}=92$ MHz for the cascade system ~\cite{Gea-Banacloche1995} and $\omega^V_{13}=2.42\times10^{9}$ MHz and $g_{12}=250$ MHz for the vee system ~\cite{PhysRevA.83.063419}, respectively. Apart from that without loss of generality, throughout the treatment the detuning value of the control laser field is set to zero.
\par
For the lambda system we choose the probe field Rabi frequency to be $g_{13}=0.5$ MHz with the decay constants $\Gamma_{31}=0.1$ MHz and $\Gamma_{32}=6$ MHz, respectively ~\cite{PhysRevA.51.R2703}. The dispersion, absorption and the group velocity profiles of the probe field are shown in Fig.2 and the resonant group velocity is found to be $v^\Lambda_g (0) \approx 17543.7$ nms$^{-1}$. To understand the population dynamics of the lambda system, in Fig.3 we illustrate the plot of the population of three levels. Here we note that at resonance ($\Delta^{\Lambda}_{31}=0$), the population of the atoms is trapped in the lowest level keeping the middle and upper two levels virtually uninhabited. This shows that in the dressed atom scenario, the lowest state of the lambda system is the dark state and this result was pointed out by many authors~\cite{Marangos1996,Zubairy1997,Fleischhauer2005}.
\begin{figure}[h]
\begin{center}
\rotatebox{0} {\includegraphics [width=16cm]{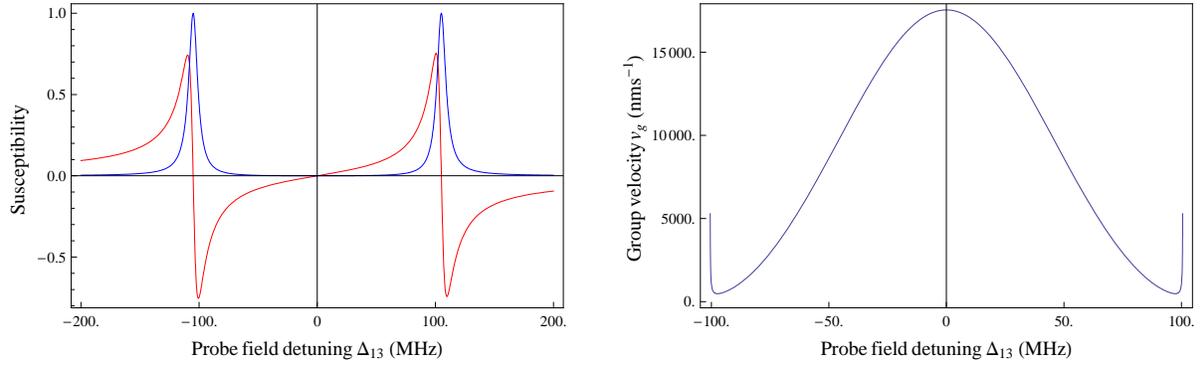}}
\end{center}
\caption{The plots of Eqs.(12) give the dispersion (Red), absorption (Blue) and the group velocity profiles of the lambda configuration as a function of the probe detuning with aforesaid values of the Rabi frequencies and decay constants
\label{graph3}}
\end{figure}
\begin{figure}[h]
\begin{center}
\rotatebox{0} {\includegraphics [width=16cm]{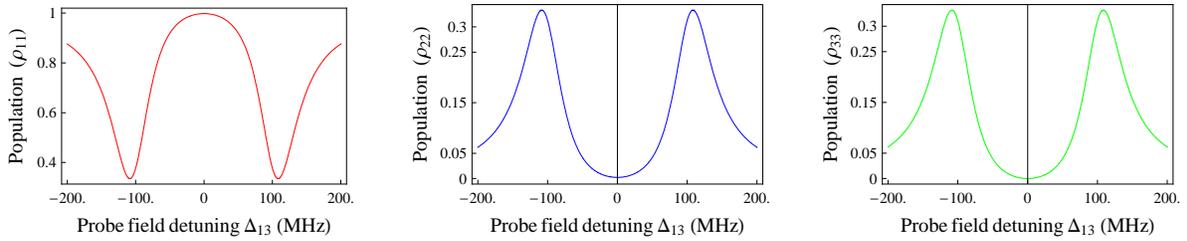}}
\end{center}
\caption{The population of the lambda system for lower (Red), middle (Blue) and upper (Green) levels from Eq.(11a) as the function of probe detuning frequency with same numerical values. We note that at resonance $\Delta_{13}=0$, all population is trapped into the lower (Red) level ($\rho_{11}\approx1$) leaving the middle (Blue) and upper (Green) levels empty ($\rho_{22}\simeq\rho_{33}\simeq0$) \label{graph3}}
\end{figure}
\par
The variation of the refractive index, coefficient of absorption and the group velocity of the cascade system with probe field Rabi frequency
$g_{12}=0.8$ MHz and decay constants $\Gamma_{21}=0.49$ MHz and $\Gamma_{32}=3.49$ MHz  ~\cite{Gea-Banacloche1995}, respectively, are depicted in Fig.4
\begin{figure}[h]
\begin{center}
\rotatebox{0} {\includegraphics [width=16cm]{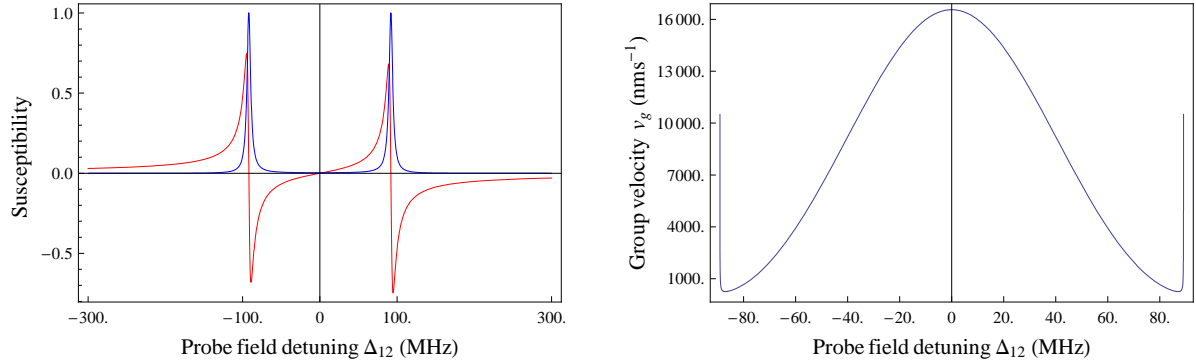}}
\end{center}
\caption{The plots give the dispersion (Red), absorption (Blue) and the group velocity profiles of the cascade system from Eqs.(16)
\label{graph5}}
\end{figure}
\begin{figure}[h]
\begin{center}
\rotatebox{0} {\includegraphics [width=16cm]{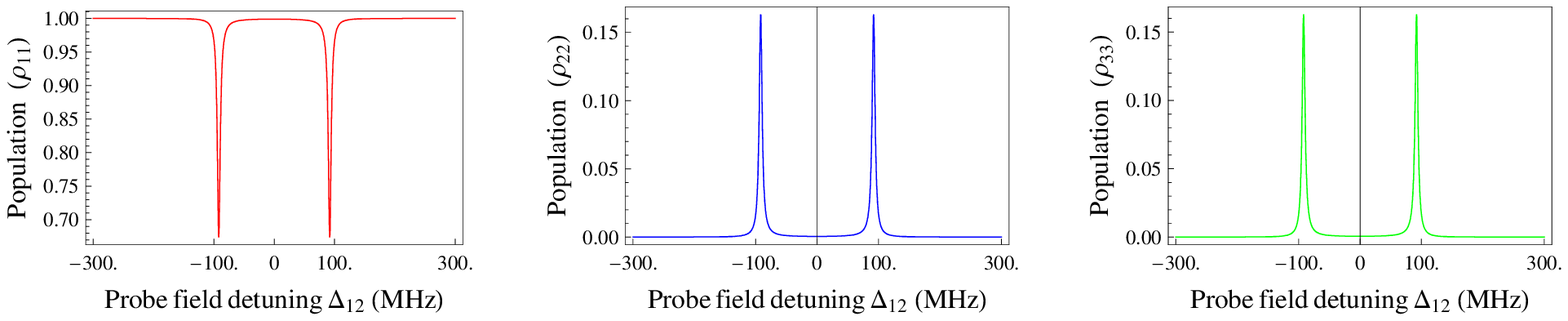}}
\end{center}
\caption{The population curves of cascade system is plotted as the function of probe field detuning frequency from Eqs(19a). We note that the population behaves similar to that of the lambda system\label{graph6}}
\end{figure}
and the group velocity of the resonant probe field is found to be $v^\Xi_{13}(0) \approx 16316.5$nms$^{-1}$. The population plot of this system is shown in Fig.5 from which it is evident that, similar to the lambda system, at resonance whole population is being accumulated into the lowest level making it a dark state leaving the middle and upper levels almost empty.
\par
Finally the dispersion, absorption and group velocity profiles for the vee system are depicted in Fig.6 with $g_{13}=10$ MHz, $\Gamma_{31}=6$ MHz and  $\Gamma_{21}=9$ MHz, respectively~\cite{PhysRevA.83.063419}, which yields the resonant group velocity to be $v^V_{13}(0)\approx$ 16558 nms$^{-1}$.
\begin{figure}[h]
\begin{center}
\rotatebox{0} {\includegraphics [width=16cm]{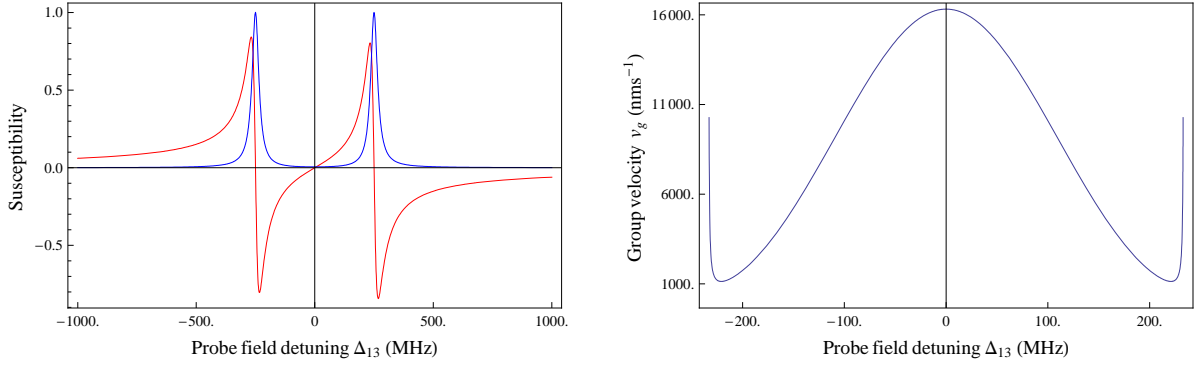}}
\end{center}
\caption{The plots of Eqs.(20) give the refractive index (Red), absorption coefficient (Blue) and the group velocity of the vee configuration
\label{graph3}}
\end{figure}
\begin{figure}[h]
\begin{center}
\rotatebox{0} {\includegraphics [width=16cm]{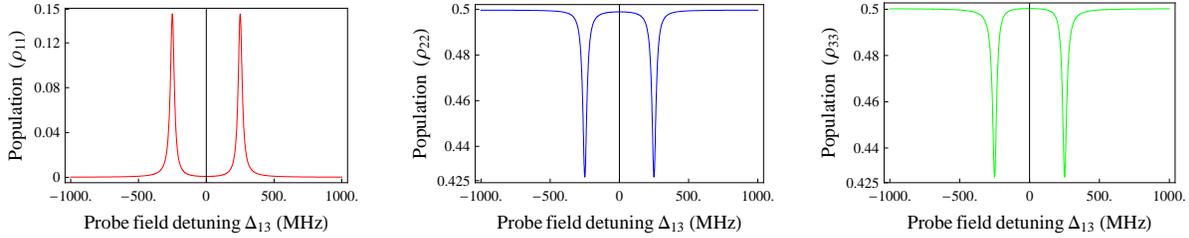}}
\end{center}
\caption{The population curves for the lower (Red), middle (Blue) and upper (Green) levels of the vee system are plotted as the function of probe field detuning frequency $\Delta_{13}$ from Eq.(19a). We note that in contrast to lambda and cascade systems, at resonance both the middle and the upper levels ($\rho_{33}\neq0, \rho_{22}\neq0$) are well populated while the lower level is almost empty ($\rho_{11}\approx 0$) \label{graph4}}
\end{figure}
Here we note that, contrast to other two configurations, a fundamental difference appears in the population dynamics of the vee system.
Fig.6 shows that when EIT occurs at probe field resonance, the atomic population is being trapped in the upper and middle levels leaving the population of the lower level almost vanishing. The reason for such unusual behaviour contrary to other two systems can be understood by comparing the dark states of three systems.
\par
The generic dark state of the three-level systems, each of which is indeed the destructive superposition of its two mutually uncoupled bare states, are given by~\cite{Fleischhauer2005},
\begin{subequations}
\begin{eqnarray}
|a_0^{\Lambda}> &=& \cos{{\theta}^{\Lambda}}|1>-\sin{{\theta}^{\Lambda}}|2>,\\
|a_0^{\Xi}> &=& \cos{{\theta}^{\Xi}}|1>-\sin{{\theta}^{\Xi}}|3>,\\
|a_0^{V}> &=& \cos{{\theta}^{V}}|2>-\sin{{\theta}^{V}}|3>,
\end{eqnarray}
\end{subequations}
for the lambda, cascade and vee systems, respectively, where $\theta^{A}$ be the mixing angle of the bare states. To get an estimate of the mixing angle we analyse the population distribution in the bare states discussed above. For the lambda system, Fig.3 shows that the accumulation of the population occurs in the lowest level (i.e., level $|1>$) and it corresponds to $\theta^{\Lambda} \simeq 0$ in Eq.(21a). Similar situation occurs for the cascade system for which in Eq.(21b) we have $\theta^{\Xi} \simeq 0$. On the other hand, since the population in the vee system is equally distributed in two bare states $|2>$ and $|3>$ as evident from Fig.7, the corresponding mixing angles in Eq.(21c) are given by $\theta^{V}\approx \frac{\pi}{4}$, respectively. Thus the dark states of the three systems in Eq.(23) are given by,
\begin{subequations}
\begin{eqnarray}
|a_0^{\Lambda}> &\simeq& |1>,\\
|a_0^{\Xi}> & \simeq & |1>,\\
|a_0^{V}> & \simeq & \frac{1}{\sqrt{2}}|2>-\frac{1}{\sqrt{2}}|3>,
\end{eqnarray}
\end{subequations}
respectively, where the maximal superposition of the bare states in the vee system is clearly evident.
\section{\label{1y} Conclusions}
This paper reports a systematic study of the EIT exhibited by the lambda, cascade and vee configurations which per se are different from one another ~\cite{Sen2012}. The Optical Bloch Equation (OBE) for each system is solved to find the steady state density matrix of the desired system which eventually gives the dispersion profile, absorption coefficient and the group velocity of resonant probe field. Although all three configurations exhibit the EIT effect, but the analysis of the population curves of each system shows that the vee system is remarkably distinct from that of the lambda and cascade systems. At resonance, whole population of the lambda and cascade systems are found to be trapped in the lowermost bare state, which is indeed effectively be the dressed dark state of these systems. On the other hand, for the vee system, the dressed dark state retains its superposed character where population is equally distributed in the middle and upper states. The estimate of the mixing angle $\theta^{A}(g_{ij},\Gamma_{ij})$ presented here is based on the numerical values of the Rabi frequencies and the decay constants when the EIT is observed at zero probe detuning. Finding the analytical expression of the angle, which determines the exact contribution of two bare states in the dark state, is possibly an open issue.
\par
Finally a few comments about the choice of the numerical values used in this paper. Since our model is generic in nature, therefore, it is easy to validate it for other values of the numerical parameters too. In particular, we have made a numerical check of our models with the parameters used in a EIT based model of micro-switching where the controlling of the windows of the Autler-Townes doublet plays key role~\cite{JianLi2011, JianLi2012}. The EIT has wide prospect of application in fabricating future quantum control devices which are capable of micro-switching, storing and processing the optical information in a micron-size cavity~\cite{Turukhin2002, Phillips2001, Liu2001}. We expect that, the maximally superposed charcater of dark state of the vee system reported here may be useful to design some advanced optically controlled switching system.
\section*{Acknowledgement}
SS is thankful to Department of Science and Technology, New Delhi for partial support. SS and TKD also thank S N Bose National Centre for Basic Sciences, Kolkata, for supporting their visit.
\bibliography{eitbib}
\vfill
\pagebreak
\appendices
\section{}
For completeness, in this Appendix we shall give the elements of the steady state density matrix of three configurations obtained by solving Optical Bloch Equation. It is worth mentioning that although the OBE can be solved without taking the control field detuning to be zero, but for mathematical convenience we work with its vanishing value for all three-level systems. (We drop the symbols $\Lambda, \Xi$ and $V$ appearing in the superscript of $\Gamma_{ij}$ and $\Delta_{ij}$ for the notational brevity.)
\begin{flushleft}
\bf{a) Lambda system:}
\end{flushleft}
\par
The denominator of Eqs.(13) for the lambda configuration is given by
\begin{eqnarray}
 \text{D}^{\Lambda} &=& \Gamma _{32} g_{13}^6+g_{23}^2 \left(\Gamma _{31}+2 \Gamma _{32}\right)
   g_{13}^4+\left(\left(2 \Gamma _{31}+\Gamma _{32}\right)
   g_{23}^4+\left(\Gamma _{31}+\Gamma _{32}\right)      \left(2
   g_{23}^2+\Gamma _{32} \left(\Gamma _{31}+\Gamma _{32}\right)\right)
   \Delta _{13}^2\right) g_{13}^2   \nonumber \\ & &     +g_{23}^2 \Gamma _{31}
   \left(g_{23}^4-2 \Delta _{13}^2 g_{23}^2+\Delta _{13}^4+\left(\Gamma
   _{31}+\Gamma _{32}\right){}^2 \Delta _{13}^2\right)
 \end{eqnarray}
while the numerators appearing in the density matrix are,
 \begin{subequations}
 \begin{eqnarray}
\rho_{\text{N}11}^{\Lambda} &=&
 g_{23}^2 \left(\Gamma _{31} \Delta _{13}^4+\Gamma _{31}
   \left(g_{13}^2-2 g_{23}^2+\left(\Gamma _{31}+\Gamma
   _{32}\right){}^2\right) \Delta
   _{13}^2+\left(g_{13}^2+g_{23}^2\right) \left(\Gamma _{32}
   g_{13}^2+g_{23}^2 \Gamma _{31}\right)\right),\nonumber \\ \\
 \rho_{\text{N}22}^{\Lambda} &=&g_{13}^2 \left(\Gamma _{32} g_{13}^4+g_{23}^2 \left(\Gamma _{31}+\Gamma
   _{32}\right) g_{13}^2+\Gamma _{32} \left(\Gamma _{31}+\Gamma
   _{32}\right){}^2 \Delta _{13}^2+g_{23}^2 \Gamma _{32} \Delta
   _{13}^2+g_{23}^4 \Gamma _{31}\right),\\
 \rho_{\text{N}33}^{\Lambda} &=&g_{13}^2 g_{23}^2 \left(\Gamma _{31}+\Gamma _{32}\right) \Delta _{13}^2,\\
 \rho_{\text{N}12}^{\Lambda} &=& -g_{13} g_{23} \left(\Gamma _{32} g_{13}^4+\left(\Gamma _{31}+\Gamma
   _{32}\right) \left(g_{23}^2+i \Gamma _{32} \Delta _{13}\right)
   g_{13}^2+g_{23}^2 \Gamma _{31} \left(g_{23}^2+i \left(\Gamma
   _{31}+\Gamma _{32}+i \Delta _{13}\right) \Delta _{13}\right)\right), \nonumber\\ \\
 \rho_{\text{N}13}^{\Lambda} &=&g_{13} g_{23}^2 \Delta _{13} \left(\Gamma _{32} g_{13}^2+g_{23}^2
   \Gamma _{31}+i \Gamma _{31} \left(\Gamma _{31}+\Gamma _{32}+i \Delta
   _{13}\right) \Delta _{13}\right),\\
 \rho_{\text{N}23}^{\Lambda} &=&-g_{13}^2 g_{23} \Delta _{13} \left(\Gamma _{31} g_{23}^2+\Gamma _{32}
   \left(g_{13}^2-i \left(\Gamma _{31}+\Gamma _{32}\right) \Delta
   _{13}\right)\right).
    \end{eqnarray}
\end{subequations}

\vfill
\pagebreak
\begin{flushleft}
\bf{b) Cascade system:}
\end{flushleft}
For the Cascade system, the denominator appearing in Eq.(17) is given by
\begin{eqnarray}
\text{D}^{\Xi} &=&
 2 \Gamma _{21} \Gamma _{32} g_{12}^6+\left\{\left(\Gamma _{21}^2+4
   \Gamma _{32} \Gamma _{21}+2 \Gamma _{32}^2\right) g_{23}^2+\Gamma
   _{21} \Gamma _{32} \left[\left(\Gamma _{21}+2 \Gamma
   _{32}\right){}^2+\Delta _{12}^2\right]\right\} g_{12}^4   \nonumber \\ &&    +\left[\left(2
   \Gamma _{21}^2+3 \Gamma _{32} \Gamma _{21}+2 \Gamma _{32}^2\right)
   g_{23}^4+\left\{\left(2 \Gamma _{21}^2+4 \Gamma _{32} \Gamma
   _{21}+\Gamma _{32}^2\right) \Delta _{12}^2  \right. \right.  \nonumber \\ &&    \left. \left.  +\Gamma _{32} \left(4
   \Gamma _{21}^3+7 \Gamma _{32} \Gamma _{21}^2+6 \Gamma _{32}^2 \Gamma
   _{21}+2 \Gamma _{32}^3\right)\right\} g_{23}^2+2 \Gamma _{21} \Gamma
   _{32} \left(\Gamma _{21}+\Gamma _{32}\right) \left\{\left(\Gamma
   _{21}+2 \Gamma _{32}\right) \Delta _{12}^2 \right. \right.   \nonumber \\ &&    \left. \left.    +\Gamma _{32} \left(\Gamma
   _{21}^2+\Gamma _{32} \Gamma _{21}+\Gamma
   _{32}^2\right)\right\}\right] g_{12}^2+\Gamma _{21} \left(\Gamma
   _{21}+\Gamma _{32}\right) \left\{g_{23}^2+\Gamma _{32} \left(\Gamma
   _{21}+\Gamma _{32}\right)\right\}\nonumber \\ &&  \left\{g_{23}^4+2 \left(\Gamma _{21}
   \Gamma _{32}-\Delta _{12}^2\right) g_{23}^2+\left(\Gamma
   _{21}^2+\Delta _{12}^2\right) \left(\Gamma _{32}^2+\Delta
   _{12}^2\right)\right\},
\end{eqnarray}
and the terms in the numerator are given by,
\begin{subequations}
\begin{eqnarray}
\rho_{\text{N}11}^{\Xi} &=&
\Gamma _{21} \Gamma _{32} g_{12}^6+\Gamma _{32} \left\{\left(\Gamma
   _{21}+\Gamma _{32}\right) g_{23}^2+\Gamma _{21} \left(\Gamma
   _{21}^2+2 \Gamma _{32} \Gamma _{21}+2 \Gamma _{32}^2+\Delta
   _{12}^2\right)\right\} g_{12}^4   \nonumber \\ &&     +\left[\Gamma _{21}^2
   g_{23}^4+\left\{\left(\Gamma _{21}^2+3 \Gamma _{32} \Gamma
   _{21}+\Gamma _{32}^2\right) \Delta _{12}^2+\Gamma _{32} \left(3
   \Gamma _{21}^3+3 \Gamma _{32} \Gamma _{21}^2+2 \Gamma _{32}^2 \Gamma
   _{21}     +\Gamma _{32}^3\right)\right\} g_{23}^2  \right.  \nonumber \\ &&  \left.  +\Gamma _{21} \Gamma_{32} \left(\Gamma _{21}+\Gamma _{32}\right) \left\{\left(\Gamma
   _{21}+3 \Gamma _{32}\right) \Delta _{12}^2+\Gamma _{32} \left(2
   \Gamma _{21}^2+\Gamma _{32} \Gamma _{21}     +\Gamma
   _{32}^2\right)\right\}\right] g_{12}^2         +\Gamma _{21} \left(\Gamma
   _{21}+\Gamma _{32}\right)  \nonumber \\ &&  \left\{g_{23}^2+\Gamma _{32} \left(\Gamma
   _{21}+\Gamma _{32}\right)\right\} \left\{g_{23}^4 +2 \left(\Gamma _{21}
   \Gamma _{32}-\Delta _{12}^2\right) g_{23}^2+\left(\Gamma
   _{21}^2+\Delta _{12}^2\right) \left(\Gamma _{32}^2+\Delta
   _{12}^2\right)\right\}, \\
\rho_{\text{N}22}^{\Xi} &=&
g_{12}^2 \left[\Gamma _{21} \Gamma _{32} g_{12}^4+\Gamma _{32}
   \left\{\left(2 \Gamma _{21}+\Gamma _{32}\right) g_{23}^2+2 \Gamma
   _{21} \Gamma _{32} \left(\Gamma _{21}+\Gamma _{32}\right)\right\}
   g_{12}^2  \right.     \nonumber \\ &&  \left.     +\left(\Gamma _{21}+\Gamma _{32}\right)
   \left\{g_{23}^2+\Gamma _{32} \left(\Gamma _{21}+\Gamma
   _{32}\right)\right\} \left\{\Gamma _{32} g_{23}^2+\Gamma _{21}
   \left(\Gamma _{32}^2+\Delta _{12}^2\right)\right\}\right], \\
\rho_{\text{N}33}^{\Xi} &=&
g_{12}^2 g_{23}^2 \left(\Gamma _{21}+\Gamma _{32}\right) \left\{\Gamma
   _{21} g_{12}^2+\left(\Gamma _{21}+\Gamma _{32}\right)
   \left(g_{23}^2+\Gamma _{21} \Gamma _{32}\right)\right\},  \\
\rho_{\text{N}12}^{\Xi} &=&
i g_{12} \left[\Gamma _{21} \Gamma _{32} \left(\Gamma _{21}+i \Delta
   _{12}\right) g_{12}^4+\Gamma _{32} \left\{\left(2 \Gamma _{21}+\Gamma
   _{32}\right) g_{23}^2+2 \Gamma _{21} \Gamma _{32} \left(\Gamma
   _{21}+\Gamma _{32}\right)\right\} \right.  \nonumber \\ &&  \left.  \left(\Gamma _{21}+i \Delta
   _{12}\right) g_{12}^2+\Gamma _{21} \left(\Gamma _{21}+\Gamma
   _{32}\right) \left\{g_{23}^2+\Gamma _{32} \left(\Gamma _{21}+\Gamma
   _{32}\right)\right\} \right.   \nonumber \\ && \left. \left\{g_{23}^2+\left(\Gamma _{21}+i \Delta
   _{12}\right)  \left(\Gamma _{32}+i \Delta _{12}\right)\right\}
   \left(\Gamma _{32}-i \Delta _{12}\right)\right],   \\
\rho_{\text{N}13}^{\Xi} &=&
g_{12} g_{23} \left[\Gamma _{21} \Gamma _{32} g_{12}^4+\left\{-\Gamma
   _{32} \Gamma _{21}^3+\Gamma _{32}^3 \Gamma _{21}+g_{23}^2
   \left(-\Gamma _{21}^2+\Gamma _{32} \Gamma _{21}+\Gamma
   _{32}^2\right)\right\} g_{12}^2   \right. \nonumber \\ && \left. -\Gamma _{21} \left(\Gamma
   _{21}+\Gamma _{32}\right) \left\{g_{23}^2+\Gamma _{32} \left(\Gamma
   _{21}+\Gamma _{32}\right)\right\} \left\{g_{23}^2+\left(\Gamma _{21}+i
   \Delta _{12}\right) \left(\Gamma _{32}+i \Delta
   _{12}\right)\right\}\right],\\
\rho_{\text{N}23}^{\Xi} &=&
i g_{12}^2 g_{23} \left(\Gamma _{21}+\Gamma _{32}\right) \left[\Gamma
   _{21} \Gamma _{32} g_{12}^2+\Gamma _{32} \left(\Gamma _{21}+\Gamma
   _{32}\right) \left(g_{23}^2+\Gamma _{21} \Gamma _{32}\right) \right. \nonumber \\ && \left.   +i
   \Gamma _{21} \left\{g_{23}^2+\Gamma _{32} \left(\Gamma _{21}+\Gamma
   _{32}\right)\right\} \Delta _{12}\right].
\end{eqnarray}
\end{subequations}
\vfill
\pagebreak
\begin{flushleft}
\bf{c) Vee system:}
\end{flushleft}
The denominator appearing in Eq.(21) is given by
\begin{eqnarray}
\text{D}^{\text{V}} &=&2 \Gamma _{21} \Gamma _{31} g_{12}^6+\left[2 \left(\Gamma
   _{21}^2+\Gamma _{31} \Gamma _{21}+\Gamma _{31}^2\right)
   g_{13}^2+\Gamma _{21} \Gamma _{31} \left\{\left(\Gamma _{21}+2 \Gamma
   _{31}\right){}^2-4 \Delta _{13}^2\right\}\right] g_{12}^4  \nonumber \\ & & +\left[2
   \Gamma _{21} \Gamma _{31} \Delta _{13}^4+\left\{\left(\Gamma
   _{21}^2+6 \Gamma _{31} \Gamma _{21}+2 \Gamma _{31}^2\right)
   g_{13}^2+4 \Gamma _{21} \Gamma _{31}^2 \left(\Gamma _{21}+\Gamma
   _{31}\right)\right\} \Delta _{13}^2  \right.   \nonumber \\ & &  \left.    +2 \left(g_{13}^2+\Gamma
   _{31}^2\right) \left(\Gamma _{21}^2+\Gamma _{31} \Gamma _{21}+\Gamma
   _{31}^2\right) \left\{g_{13}^2+\Gamma _{21} \left(\Gamma _{21}+\Gamma
   _{31}\right)\right\}\right] g_{12}^2  \nonumber \\ & &   +\Gamma _{21} \Gamma _{31}
   \left(2 g_{13}^2+\Gamma _{31}^2+\Delta _{13}^2\right)
   \left\{\left(g_{13}^2+\Gamma _{21} \left(\Gamma _{21}+\Gamma
   _{31}\right)\right){}^2+\Gamma _{21}^2 \Delta _{13}^2\right\},
\end{eqnarray}
and other terms in the numerator of the density matrix are,
\begin{subequations}
\begin{eqnarray}
\rho_{\text{N}11}^{\text{V}} &=& \Gamma _{21} \Gamma _{31} g_{12}^6+\left\{\left(\Gamma _{21}^2+\Gamma
   _{31} \Gamma _{21}+\Gamma _{31}^2\right) g_{13}^2+\Gamma _{21}
   \Gamma _{31} \left(\Gamma _{21}^2+2 \Gamma _{31} \Gamma _{21}+2
   \Gamma _{31}^2-2 \Delta _{13}^2\right)\right\}
   g_{12}^4  \nonumber \\ & &   +\left[\left(\Gamma _{21}^2+\Gamma _{31} \Gamma _{21}+\Gamma
   _{31}^2\right) g_{13}^4+\left\{\Gamma _{21}^4+2 \Gamma _{31} \Gamma
   _{21}^3+4 \Gamma _{31}^2 \Gamma _{21}^2+2 \Gamma _{31}^3 \Gamma
   _{21}+\Gamma _{31}^4  \right. \right.   \nonumber \\ & & \left. \left. +\Gamma _{31} \left(2 \Gamma _{21}+\Gamma
   _{31}\right) \Delta _{13}^2\right\} g_{13}^2+\Gamma _{21} \Gamma
   _{31} \left\{2 \Gamma _{31} \Gamma _{21}^3+\left(3 \Gamma
   _{31}^2-\Delta _{13}^2\right) \Gamma _{21}^2  \right. \right.  \nonumber \\ & & \left. \left. +2 \Gamma _{31}
   \left(\Gamma _{31}^2+\Delta _{13}^2\right) \Gamma _{21}+\left(\Gamma
   _{31}^2+\Delta _{13}^2\right){}^2\right\}\right] g_{12}^2+\Gamma
   _{21} \Gamma _{31} \left(g_{13}^2+\Gamma _{31}^2+\Delta
   _{13}^2\right) \nonumber \\ & &  \left\{\left(g_{13}^2+\Gamma _{21} \left(\Gamma
   _{21}+\Gamma _{31}\right)\right){}^2+\Gamma _{21}^2 \Delta
   _{13}^2\right\},  \\
\rho_{\text{N}22}^{\text{V}} &=&g_{12}^2 \left[\Gamma _{21} \Gamma _{31} g_{12}^4+\left\{\left(\Gamma
   _{21}^2+\Gamma _{31}^2\right) g_{13}^2+2 \Gamma _{21} \Gamma _{31}
   \left(\Gamma _{31} \left(\Gamma _{21}+\Gamma _{31}\right)-\Delta
   _{13}^2\right)\right\} g_{12}^2  \right.    \nonumber \\ & &   \left.   +\Gamma _{31} \left\{\Gamma _{21}
   g_{13}^4+\left[\Gamma _{21}^3+\Gamma _{31} \Gamma _{21}^2+\Gamma
   _{31}^2 \Gamma _{21}+\Gamma _{31}^3+\left(3 \Gamma _{21}+\Gamma
   _{31}\right) \Delta _{13}^2\right] g_{13}^2  \right. \right.    \nonumber \\ & &  \left. \left.  +\Gamma _{21}
   \left(\Gamma _{31}^2+\Delta _{13}^2\right) \left[\left(\Gamma
   _{21}+\Gamma _{31}\right){}^2+\Delta _{13}^2\right]\right\}\right],\\
\rho_{\text{N}33}^{\text{V}} &=&g_{13}^2 \left[\Gamma _{21} \Gamma _{31} g_{12}^4+\left\{\left(\Gamma
   _{21}^2+\Gamma _{31}^2\right) g_{13}^2+\Gamma _{21} \left(\Gamma
   _{21}+\Gamma _{31}\right) \left(\Gamma _{21}^2+\Gamma _{31}^2+\Delta
   _{13}^2\right)\right\} g_{12}^2  \right.  \nonumber \\ & &     \left. +\Gamma _{21} \Gamma _{31}
   \left\{\left(g_{13}^2+\Gamma _{21} \left(\Gamma _{21}+\Gamma
   _{31}\right)\right){}^2+\Gamma _{21}^2 \Delta _{13}^2\right\}\right],  \\
\rho_{\text{N}12}^{\text{V}} &=&i g_{12} \left[\Gamma _{21}^2 \Gamma _{31} g_{12}^4+\left\{\left(\Gamma
   _{21}^3+\Gamma _{31}^2 \Gamma _{21}+2 i \Gamma _{31}^2 \Delta
   _{13}\right) g_{13}^2     +2 \Gamma _{21}^2 \Gamma _{31} \left[\Gamma
   _{31} \left(\Gamma _{21}+\Gamma _{31}\right)  \right. \right. \right. \nonumber \\ & &  \left. \left. \left.   -\Delta
   _{13}^2\right]\right\} g_{12}^2+\Gamma _{21} \Gamma _{31}
   \left\{g_{13}^2+\Gamma _{21} \left(\Gamma _{21}+\Gamma _{31}-i \Delta
   _{13}\right)\right\} \left\{\left(\Gamma _{21}+2 i \Delta _{13}\right)
   g_{13}^2  \right. \right.  \nonumber \\ && \left. \left.  +\left(\Gamma _{21}+\Gamma _{31}+i \Delta _{13}\right)
   \left(\Gamma _{31}^2+\Delta _{13}^2\right)\right\}\right],  \\
 \rho_{\text{N}13}^{\text{V}} &=&
i g_{13} \left[\Gamma _{21} \Gamma _{31} \left(\Gamma _{31}-i \Delta
   _{13}\right) g_{12}^4+\left\{i \Gamma _{21} \Gamma _{31} \Delta
   _{13}^3+\Gamma _{21} \Gamma _{31} \left(\Gamma _{21}+\Gamma
   _{31}\right) \Delta _{13}^2  \right. \right.  \nonumber \\ &&  \left. \left. +i \left(g_{13}^2+\Gamma _{21} \Gamma
   _{31}\right) \left(\Gamma _{31}^2-\Gamma _{21}^2\right) \Delta
   _{13}+\Gamma _{31} \left(\Gamma _{21}^2+\Gamma _{31}^2\right)
   \left[g_{13}^2+\Gamma _{21} \left(\Gamma _{21}+\Gamma
   _{31}\right)\right]\right\} g_{12}^2  \right. \nonumber \\ &&  \left. +\Gamma _{21} \Gamma _{31}
   \left(\Gamma _{31}+i \Delta _{13}\right) \left\{\left[g_{13}^2+\Gamma
   _{21} \left(\Gamma _{21}+\Gamma _{31}\right)\right]{}^2+\Gamma
   _{21}^2 \Delta _{13}^2\right\}\right],    \\
\rho_{\text{N}23}^{\text{V}} &=&
g_{12} g_{13} \left[\Gamma _{21} \Gamma _{31}
   g_{12}^4+\left\{\left(\Gamma _{21}^2+\Gamma _{31}^2\right)
   g_{13}^2+\Gamma _{21} \Gamma _{31} \left[\Gamma _{21}^2+2 \Gamma
   _{31} \Gamma _{21}+\left(\Gamma _{31}+i \Delta
   _{13}\right){}^2\right]\right\} g_{12}^2 \right. \nonumber \\ && \left.+\Gamma _{21} \Gamma _{31}
   \left\{g_{13}^2+\Gamma _{21} \left(\Gamma _{21}+\Gamma _{31}+i \Delta
   _{13}\right)\right\} \left\{g_{13}^2+\Delta _{13}^2+\Gamma _{31}
   \left(\Gamma _{21}+\Gamma _{31}\right)+i \Gamma _{21} \Delta
   _{13}\right\}\right]. \nonumber \\
\end{eqnarray}
\end{subequations}

\end{document}